\documentclass[prb,twocolumn,showpacs,amsmath,amssymb,superscriptaddress]{revtex4-1}
\usepackage{graphicx}
\usepackage{color}
\usepackage{braket}
\usepackage{amsmath}
\usepackage{amsfonts}

\usepackage{amsmath,amssymb,mathrsfs}
\usepackage{bbm}
\usepackage[dvipsnames]{xcolor}

\date{\today}

\begin{document}
	
	\title{Fractional boundary charges with quantized slopes in interacting \\ one- and two-dimensional systems}
	
	\author{Katharina Laubscher}
	\affiliation{Department of Physics, University of Basel, Klingelbergstrasse 82, 
		CH-4056 Basel, Switzerland}
	
	\author{Clara S. Weber}
	\affiliation{Institut f\"ur Theorie der Statistischen Physik, RWTH Aachen, 
		52056 Aachen, Germany and JARA - Fundamentals of Future Information Technology}
	
	\author{Dante M. Kennes}
	\affiliation{Institut f\"ur Theorie der Statistischen Physik, RWTH Aachen, 
		52056 Aachen, Germany and JARA - Fundamentals of Future Information Technology}
	\affiliation{Max Planck Institute for the Structure and Dynamics of Matter, Center for Free Electron Laser Science, 22761 Hamburg, Germany}
	
	\author{Mikhail Pletyukhov}
	\affiliation{Institut f\"ur Theorie der Statistischen Physik, RWTH Aachen, 
		52056 Aachen, Germany and JARA - Fundamentals of Future Information Technology}
	
	\author{Herbert Schoeller}
	\affiliation{Institut f\"ur Theorie der Statistischen Physik, RWTH Aachen, 
		52056 Aachen, Germany and JARA - Fundamentals of Future Information Technology}
	
	\author{Daniel Loss}
	\affiliation{Department of Physics, University of Basel, Klingelbergstrasse 82, 
		CH-4056 Basel, Switzerland}
	
	\author{Jelena Klinovaja}
	\affiliation{Department of Physics, University of Basel, Klingelbergstrasse 82, 
		CH-4056 Basel, Switzerland}
	
	\begin{abstract}
		We study fractional boundary charges (FBCs) for two classes of strongly interacting systems. First, we study strongly interacting nanowires subjected to a periodic potential with a period that is a rational fraction of the Fermi wavelength. For sufficiently strong interactions, the periodic potential leads to the opening of a charge density wave gap at the Fermi level. The FBC then depends linearly on the phase offset of the potential with a quantized slope determined by the period. Furthermore, different possible values for the FBC at a fixed phase offset label different degenerate ground states of the system that cannot be connected adiabatically. Next, we turn to the fractional quantum Hall effect (FQHE) at odd filling factors $\nu=1/(2l+1)$, where $l$ is an integer. For a Corbino disk threaded by an external flux, we find that the FBC depends linearly on the flux with a quantized slope that is determined by the filling factor. Again, the FBC has $2l+1$ different branches that cannot be connected adiabatically,
		reflecting the $(2l+1)$-fold  degeneracy of the ground state. These results allow for several promising and strikingly simple ways to probe strongly interacting  phases via boundary charge measurements.
	\end{abstract}
	
	\maketitle
	
	\textit{Introduction.} %
	The emergence of fractional charges in topologically nontrivial systems is a recurring theme in modern condensed matter physics that has been discussed in several different contexts. In the fractional quantum Hall effect (FQHE), for example, strong electron-electron interactions lead to the emergence of exotic quasiparticles carrying only a fraction of the electronic charge $e$.~\cite{Moore1991,Willett1987,Halperin1987,Haldane1983} On the other hand, well-defined fractional charges can also emerge in the ground state of topological insulators. Early examples include the Jackiw-Rebbi~\cite{Jackiw1976,Jackiw1981} and Su-Schrieffer-Heeger~\cite{Su1979,Su1981} models, where domain walls between topologically non-equivalent phases bind fractional charges that are quantized due to symmetry. Generally, fractional boundary charges (FBCs) can accumulate at the boundaries of an insulator. Importantly, the possible presence of edge states influences the total boundary charge only by an integer number, while the \emph{fractional} part of the boundary charge contains contributions from all extended states and is directly related to bulk properties via the Zak-Berry phase.~\cite{vanderbilt_kingsmith_prb_93,vanderbilt_book_18,Resta1993,Resta1994,Ortiz1994,Rhim2017,Pletyukhov2020b} FBCs of this type have been studied in a large variety of systems, including different types of one-dimensional (1D) models,~\cite{Rice1982,Heeger1988,Kivelson1983,Jackiw1983,Qi2008,Vayrinen2011,Klinovaja2012,Klinovaja2013,Rainis2014,Sticlet2014,Wakatsuki2014,Klinovaja2015,Park2016,Fleckenstein2016,Lil,Thakurathi2018,Jeong2019,Jana2019,Pletyukhov2020a,Pletyukhov2020b,Pletyukhov2020c,Lin2020,Yang2020,Weber2020,Lin2021} topological crystalline insulators,~\cite{Hughes2011,Lau2016,Alexandradinata2016,Miert2017,Lau2018} higher-order topological insulators,~\cite{Benalcazar2017a,Benalcazar2017b,Miert2018,Benalcazar2019,Peterson2020,Watanabe2020,Hirosawa2020,Takahashi2021} and the integer quantum Hall effect (IQHE).~\cite{Thakurathi2018} While the presence of symmetries leads to a quantization of the FBC in rational units,\cite{Pletyukhov2020c} certain \emph{universal} features of the FBC persist even in the absence of symmetries. For generic 1D tight-binding models with periodically modulated on-site potentials, it was shown that the FBC is a sharp quantity~\cite{note5} that changes linearly with the phase offset of the modulation with a universally quantized slope even in the presence of disorder.\cite{Park2016,Thakurathi2018,Pletyukhov2020a,Pletyukhov2020b} Furthermore, this slope can be directly related to the Hall conductance in the 2D IQHE.~\cite{Thakurathi2018}
	
	\begin{figure}[bt]
		\centering
		\includegraphics[width=0.8\columnwidth]{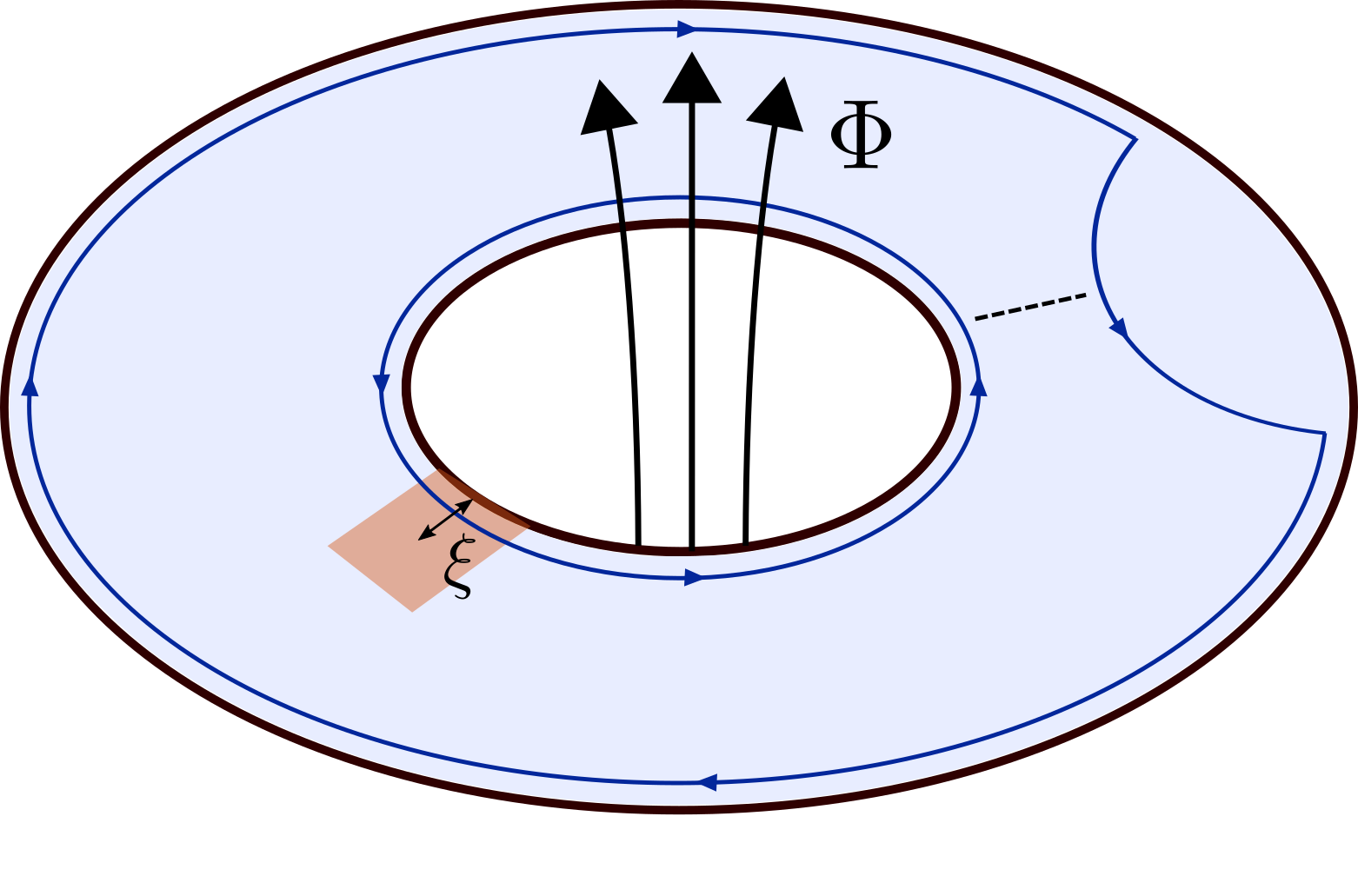}
		\caption{Corbino disk in the FQHE regime threaded by an external flux $\Phi$. The FBC is measured in the 
			red region, which extends into the bulk on the order of a few edge state localization lengths $\xi$. In the presence of a constriction, indicated by the dashed line, tunneling of fractional charges between the chiral edge states (blue lines) is allowed.}
		\label{fig:corbino}
	\end{figure} 
	
	Motivated by these results on noninteracting systems, the aim of this work is to study the universal properties of the FBC in strongly interacting systems. First, we consider a 1D nanowire with a periodic potential of the form $V_m(x)=2V_m\cos(2mk_Fx+\alpha)$, where $k_F$ is the Fermi momentum, $m$ an integer, and $\alpha$ a phase offset. For $m=1$, it is well-known that a charge density wave (CDW) gap is opened at the Fermi level~\cite{Gangadharaiah2012} with an FBC that depends linearly on $\alpha$ with a quantized slope $1/2\pi$.~\cite{Park2016,Thakurathi2018} In the presence of strong interactions, additional gaps can be opened for $m>1$. We show that in this case the FBC depends again linearly on $\alpha$ with a \emph{fractional} slope $1/2\pi m$. Perhaps even more interestingly, we find that there are now $m$ degenerate ground states labeled by $m$ different branches of the FBC that cannot be connected under adiabatic evolution of $\alpha$.
	
	Next, we extend our considerations to a two-dimensional electron gas (2DEG) in the FQHE regime at odd filling factors $\nu=1/(2l+1)$, where $l$ is an integer. For a Corbino disk threaded by an external flux $\Phi$ as shown in Fig.~\ref{fig:corbino}, we find that the FBC depends linearly on $\Phi$ with a slope determined by the filling factor. From this the standard Hall conductance follows. Further, the FBC has $2l+1$ distinct branches labeling $2l+1$ degenerate ground states that cannot be adiabatically connected unless fractional charges are allowed to tunnel between the two boundaries due to, e.g., a constriction, see again Fig.~\ref{fig:corbino}. We outline how these results open up a strikingly simple way to probe strongly interacting systems via boundary charge measurements.
	
	\textit{FBC in one dimension.} %
	We first study the FBC in a 1D nanowire of spinless electrons subjected to a spatially modulated potential $V_m(x)$.\cite{note1}
	The single-particle Hamiltonian reads
	\begin{equation}
	H=\int dx\ \Psi^\dagger(x)\left[-\frac{\hbar^2\partial_x^2}{2m^*}+V_m(x)\right]\Psi(x),
	\end{equation}
	where $\Psi^\dagger(x)$ [$\Psi(x)$] creates [annihilates] a spinless electron of mass $m^*$ at the position $x$. Without interactions, the onsite potential opens a CDW gap at the Fermi level only for $m=1$.~\cite{Gangadharaiah2012} To study the more general case of an interacting system, we linearize the spectrum around the Fermi points and write $\Psi(x)=R(x)e^{ik_Fx}+L(x)e^{-ik_Fx}$, where $R(x)$ and $L(x)$ are slowly varying right- and left-moving fields, respectively. Next, we introduce chiral bosonic fields $R(x)\propto e^{i\phi_1(x)}$ and $L(x)\propto e^{i\phi_{\bar 1}(x)}$ satisfying standard commutation relations~\cite{Giamarchi2004} $[\phi_r(x),\phi_{r'}(x')]=ir\pi\delta_{rr'}\mathrm{sgn}{(x-x')}$. This ensures the correct anticommutation relation between fermionic operators of the same species, while the remaining commutation relations can be ensured by Klein factors, which we do not include explicitly. It is also useful to define local conjugate fields $\phi=(\phi_{\bar 1}-\phi_{ 1})/2$ and $\theta=(\phi_{\bar 1}+\phi_{ 1})/2$ with $[\phi(x),\theta(x')]=\frac{i\pi}{2}\mathrm{sgn}(x-x')$. Small-momentum interactions are now included via the standard kinetic term $H_0=\frac{v}{2\pi}\int dx\,\{K[\partial_x\theta(x)]^2+\frac{1}{K}[\partial_x\phi(x)]^2\}$, where $v$ is the velocity and $K$ the Luttinger liquid parameter.~\cite{Giamarchi2004} Furthermore, momentum-conserving multi-electron processes involving backscatterings can lead to the opening of gaps when relevant in the renormalization group (RG) sense.~\cite{Giamarchi2004,Kane2002,Klinovaja2014a,Klinovaja2014b,Oreg2014,Teo2014} In our case, to lowest order in the interaction, the corresponding term reads
	\begin{equation}
	H_{CDW}^{m} = \tilde{V}_m\int dx\, [(R^\dagger L)^m e^{i\alpha}+ \mathrm{H.c.}],
	\label{eq:HCDW_interactions}
	\end{equation}
	where we have neglected rapidly oscillating contributions. Here, $\tilde{V}_m\propto V_m g_B^{m-1}$, where $g_B$ is the strength of the backscattering term induced by interactions and $V_m>0$. In terms of the bosonic fields, the CDW term takes the form $H_{CDW}^m=\int dx\,\mathcal{H}_{CDW}^m(x)$ with
	\begin{align}
	\mathcal{H}_{CDW}^m(x)=\frac{{-2|\tilde V_m|}}{(2\pi a)^m} \cos (2m\phi(x)+\alpha-\alpha_0),
	\label{eq:H_CDW_bosonic}
	\end{align}
	where $a$ is a short-distance cutoff and $\alpha_0$ an irrelevant phase shift. 
	The above term is of sine-Gordon form and opens a full gap at the Fermi level whenever relevant in the RG sense. This can be achieved if $K<2/m^2$ (which generally requires long-range interactions) or if the bare coupling constant is already of order one compared to the Fermi energy. From now on, we therefore focus on the case where $H_{CDW}^m$ is relevant.
	
	We now consider a semi-infinite system with a single boundary at $x=0$. In the semiclassical limit of infinitely strong pinning, the bosonic field $\phi$ takes a constant bulk value in order to minimize the cosine term. Explicitly, we find 
	$-\phi(\infty)=(\alpha-\alpha_0)/2m+p\pi/m$, where $p$ is an integer. At the edge of the system at $x=0$, on the other hand, we impose vanishing boundary conditions by demanding $R(0)+L(0)=0$. This implies 
	$\phi_1(0)-\phi_{\bar{1}}(0)=-2\phi(0)=\pi\ \mathrm{mod}\ 2\pi$. We then define the FBC $Q_B$ as the excess charge at the boundary of the system as compared to a constant bulk contribution. Using that the electron density is $\rho(x)=-\partial_x \phi (x)/\pi$, we have ${Q_B} = -[\phi(\infty)-\phi(0)]/\pi$ in units of the electron charge $e$. Plugging in the bulk and edge values for $\phi$ found above, we obtain {for $Q_B^{1D}\equiv Q_B$} (up to an irrelevant constant)
	\begin{equation}
	Q_B^{1D}=\frac{\alpha}{2\pi m}+\frac{{p}}{m}\quad\mathrm{mod} \,1.
	\label{eq:FBC_1D_fractional}
	\end{equation}
	This result has several interesting features: Firstly, we see that the FBC is a linear function of $\alpha$ with a slope $1/2\pi m$. For $m=1$, this agrees with the result that was previously obtained for noninteracting systems,\cite{Park2016,Thakurathi2018} but the derivation presented here also holds in the presence of interactions.\cite{note2} Secondly, for fixed $\alpha$, there are $m$ different values for the FBC, $Q_B^{1D}-\alpha/2\pi m\in\{0,1/m,...,(m-1)/m\}$. For $m>1$, we therefore find that the ground state is $m$-fold degenerate. Thirdly, these different ground states cannot be connected to one another under adiabatic evolution of $\alpha$. As such, a given branch of the FBC is $2\pi m$-periodic, while the Hamiltonian is $2\pi$-periodic. Finally, we emphasize that these results are independent of the exact value of $K$ but hold whenever $H_{CDW}^m$ is relevant.
	
	In fact, Eq.~(\ref{eq:FBC_1D_fractional}) can also be understood from more general arguments without the use of the bosonization formalism. To see this, let us assume that the bulk is fully gapped by the backscattering mechanism discussed above. If we shift the origin of the system by $\pi/mk_F\equiv \lambda_F/2m$, the FBC cannot change. Furthermore, any shift of the lattice by $d$ can always be compensated by shifting $\alpha$. Thus, the FBC is a function of both of them and necessarily has the form $Q_B (\alpha,d)\equiv Q_B(\alpha/2\pi+2md/\lambda_F)$. This is nothing but a form of `Galilean invariance' in $\alpha$ and $d$. On the other hand, a shift by $d$ changes $Q_B$ by $d\bar\rho_B$, where $\bar\rho_B=2/\lambda_F$ is the average bulk density. Thus, we find that the FBC is a linear function of not only $d$ but also $\alpha$ and has the form $Q_B = (\alpha/2\pi+2md/\lambda_F)/m + C$, where $C$ is a constant. Again, we find that the slope of the phase dependence is $1/2\pi m$. Simultaneously, there must be $m$ different branches of the FBC [corresponding to $m$ values $C=0,1/m,...,(m-1)/m$] 
	since $H$ is $2\pi$-periodic.
	
	\begin{figure}[t!]
		\centering
		\includegraphics[width =\columnwidth]{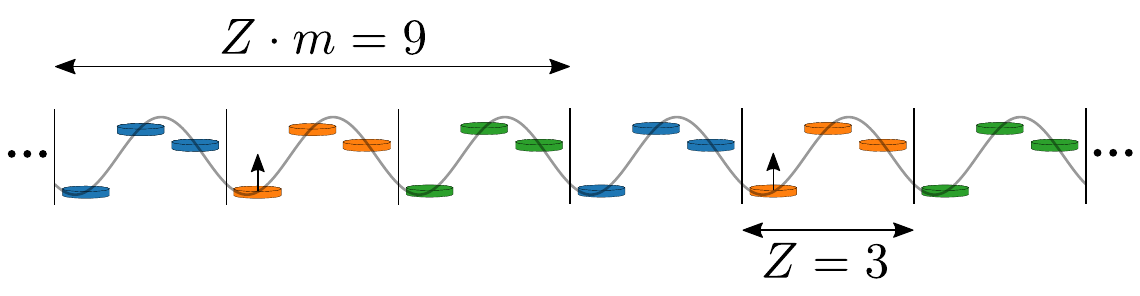}
		\caption{Pictorial representation of the ground state for $m=Z=3$ and a certain phase $\alpha$ in the investigated regime. Every $9$th site is occupied by a particle, which is indicated by the small arrows pointing up. The ground state is three-fold degenerate as one can move the particles to the green or blue unit cells. 
		}
		\label{fig:GS_eff_model}
	\end{figure}
	
	\textit{Effective model.} To illustrate the $m$-fold degenerate ground state and the phase dependence of the FBC [see Eq.~(\ref{eq:FBC_1D_fractional})], we consider the following tight-binding model with $N_s$ sites \begin{align}
	H=&- t \sum_{n=1}^{N_s-1} (a_n^\dagger a_{n+1} +\mathrm{H.c.})+ \sum_{l=1}^r \sum_{n=1}^{N_s-l} U_l \, \hat{\rho}_n \hat{\rho}_{n+l} \notag \\
	&+\sum_{n=1}^{N_s} v_{ex} \cos\left(\frac{2\pi}{Z} n +\alpha \right) \hat{\rho}_n \, ,
	\label{eq:H_effective_model}
	\end{align}
	where $\hat{\rho}_n=a^\dagger_n a_n$, $t$ is the hopping amplitude, $v_{ex}$ the amplitude of the potential modulation with period $Z$ and phase $\alpha$, and $U_l$ the electron-electron interaction with range $r$, which is required to be sufficiently long-ranged. We choose $U_l=U$ and $r=Z(m-1)$, where $\bar{\rho}_B=\frac{1}{mZ}$ is the average bulk density (corresponding to filling $\nu=\frac{1}{m}$). For sufficiently large $v_{ex}$ or $U$, this means that every $m$th minimum is occupied in the thermodynamic limit (see Fig.~\ref{fig:GS_eff_model}). The ground state is then $m$-fold degenerate because one could shift all particles simultaneously to the next minimum. 
	
	\begin{figure}[t!]
		\centering
		\includegraphics[width =\columnwidth]{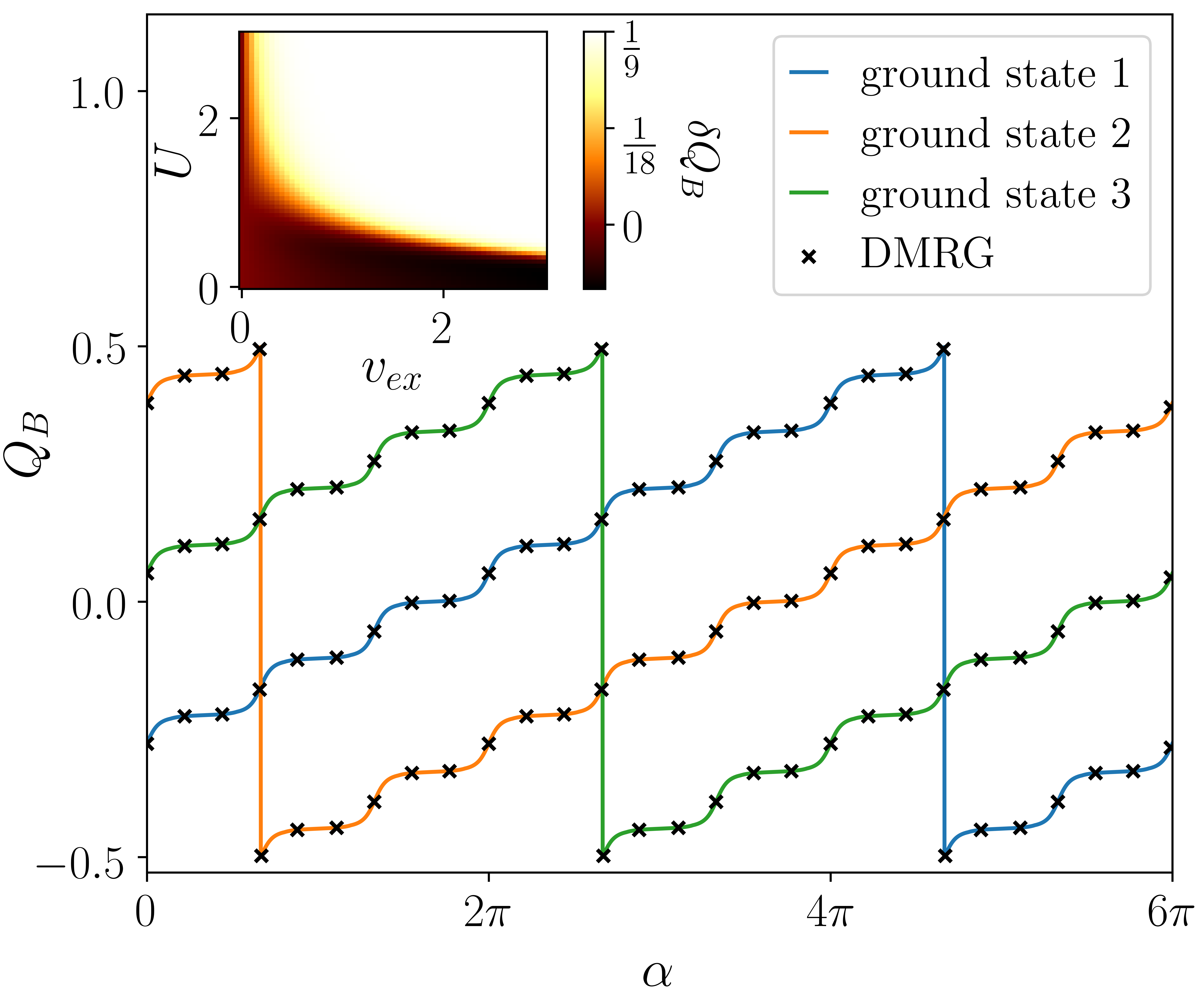}
		\caption{FBC $Q_B$ of the effective model with $m=Z=3$ and $t=1$.  
			Main panel: $Q_B$ as a function of $\alpha$ at $v_{ex}=5$ and $U=10$ calculated with DMRG (crosses) and perturbation theory (solid lines). For details about the calculation see Appendices A, B, and C. The FBC shows a linear slope $1/6\pi$ up to a periodic function and a jump of size unity. Inset: False color plot showing $\delta Q_B= Q_B \left(2\pi\right) - Q_B\left(\frac{4\pi}{3}\right)$ for different $v_{ex}$ and $U$. The expected linear slope (white region) is observed already for relatively small $v_{ex}$ and intermediate $U$.}
		\label{fig:QB_eff_model}
	\end{figure}
	
	Next, we investigate the evolution of the FBC with $\alpha$, see Fig.~\ref{fig:QB_eff_model}. We calculate $Q_B$ with perturbation theory for $U \gg v_{ex} \gg t$, see Appendix~B for details, and compare these results with results obtained from a numerically exact density matrix renormalization group (DMRG) approach. For an open system of finite size one gets a larger degeneracy of the ground state as one can also shift single particles close to the boundaries. Thus, we do not use an integer number of filling unit cells of size $Z\cdot m$ but cut some sites at the boundary. The ground state of the system is then nondegenerate and we can perform a variational ground state search. For more details we refer to Appendix~A. 
	
	We confirm, in accordance with Eq.~(\ref{eq:FBC_1D_fractional}), that 
	$Q_B$ shows a linear slope $1/2\pi m=1/6\pi$ up to a $2\pi/Z$-periodic function (with $Z=3$).\cite{note6} The inset of Fig.~\ref{fig:QB_eff_model} shows a false color plot demonstrating for which values of $v_{ex}$ and $U$ this linear slope indicated by $\delta Q_B = Q_B \left(2\pi\right) - Q_B\left(\frac{4\pi}{3}\right)=\frac{1}{9}$ (white region) is stabilized. We observe that this is already the case for relatively small $v_{ex}$ and intermediate $U$. The general phenomenology of fractionally quantized slopes in the FBC of this 1D model is therefore quite general and does not require fine tuning. 
	The additional modulation by a $2\pi/Z$-periodic function that was not present in Eq.~(\ref{eq:FBC_1D_fractional}) is a consequence of commensurability between the lattice constant and the Fermi wavelength and vanishes in the continuum limit.~\cite{Park2016} In Appendices~D and E we show that our results are stable against disorder of the hoppings and the on-site potentials and that a quantized slope is also present in the case of long ranged interactions $U_l \propto l^{-6}$ or $U_l \propto\exp(-\gamma l)\, l^{-2}$.
	
\begin{figure}[bt]
		\centering
		\includegraphics[width=0.85\columnwidth]{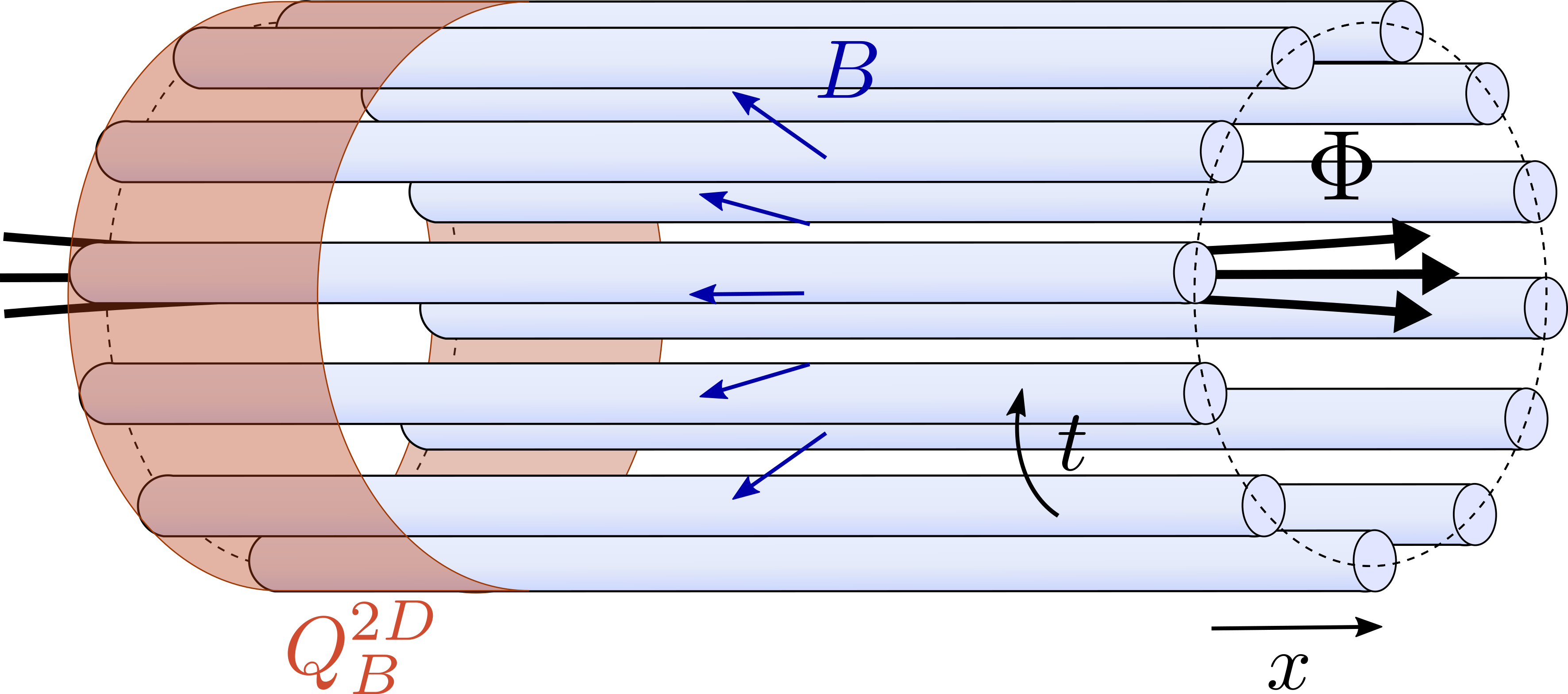}
		\caption{An array of  
			nanowires tunnel-coupled by $t$ arranged to a cylinder threaded by an external flux $\Phi$. A magnetic field $B$ perpendicular to the cylinder surface  drives the system into an FQHE phase. The FBC $Q_B^{2D}$ is calculated in the red region and shown to vary linearly with $\Phi$, see Eq.~(\ref{eq:FBC_FQHE}).
		}
		\label{fig:cylinder}
	\end{figure}

	\textit{FBC in two dimensions - FQHE.} %
	Next, we study the FBC in a 2DEG in the FQHE regime at odd filling factors $\nu=1/(2l+1)$ for an integer $l$. To facilitate the analytical treatment of strong interactions, we make use of a coupled-wire construction of the FQHE.~\cite{Kane2002,Teo2014} We consider an array of $N$ parallel nanowires, where the individual wires are oriented along the $x$ axis and the wires are stacked along the $y$ axis.\cite{Poilblanc1987,Gorkov1995} We assume periodic boundary conditions along the latter, realizing the cylinder geometry shown in Fig.~\ref{fig:cylinder}. The kinetic term is $H_0=\sum_n H_{0,n}$ with $H_{0,n} =-\frac{\hbar^2}{2m^*}  \int dx\ \Psi_{n}^\dagger(x)\,\partial_x^2\,\Psi_{n}(x)$, where $\Psi_n^\dagger(x)$ [$\Psi_n(x)$] creates [annihilates] a spinless electron of mass $m^*$ at the position $x$ in the $n$th wire. A magnetic field $\bf B$ (${\bf B'}$) is applied perpendicular to the surface (along the axis) of the cylinder and the vector potential is chosen as ${\bf A} = Bx \hat y$ [${\bf A'}=(B'R/2)\hat y$], where $R=Na_y/2\pi$ is the radius of the cylinder and $a_y$ denotes the interwire distance. Finally, the tunneling between neighboring nanowires is described by $H_T = \sum_n H_{T, n+1/2}$ with
	\begin{align}
	H_{T, n+1/2}=te^{i\varphi}\int dx\ e^{i k_B x} \Psi_{n+1}^\dagger \Psi_{n} + \mathrm{H.c.}
	\end{align}
	Here, $k_B= eB a_y/\hbar$ and $\varphi=\frac{e}{\hbar}\frac{\Phi}{N}$, with the total flux through the cylinder given by $\Phi=\pi R^2B'$. 
	To treat interactions, we again linearize the spectrum around the Fermi points, $\Psi_{n}(x)=R_n(x)e^{ik_Fx}+L_n(x)e^{-ik_Fx}$, and switch to a bosonized language by writing $R_n(x)\propto e^{i\phi_{1n}(x)}$, $L_n(x)\propto e^{i\phi_{\bar 1n}(x)}$ with $[\phi_{rn}(x),\phi_{r'n'}(x')]=ir\pi\delta_{rr'}\delta_{nn'}\mathrm{sgn}{(x-x')}$. Small-momentum interactions can then be included in the standard way and lead to a gapless sliding Luttinger liquid phase~\cite{Teo2014} that we will not characterize here. For our purposes, it suffices to note that if $k_F$ becomes commensurable with $k_B$ such that $2k_F/k_B=\nu$, an additional momentum-conserving multi-electron process can be constructed such that a gap is opened at the Fermi level and the FQHE at filling $\nu$ is realized.~\cite{Teo2014} Explicitly, the term that opens the gap is $H_{T, n+1/2}^{l}=\int dx\,\mathcal{H}_{T,n+1/2}^l$ with~\cite{note3}
	\begin{align}
	\mathcal{H}_{T, n+1/2}^{l}=t_l e^{i\varphi}R_{n+1}^\dagger L_{n} (R_{n+1}^\dagger L_{n+1})^l(R_{n}^\dagger L_{n})^l+\mathrm{H.c.}
	\label{eq:Htun_interactions}
	\end{align}
	Here, $t_{l}\propto tg_B^{2l}$. Introducing the fields 
	$\eta_{rn}=(l+1)\phi_{rn}-l\phi_{\bar{r}n}$, Eq.~(\ref{eq:Htun_interactions}) becomes
	\begin{equation}
	\mathcal{H}_{T, n+1/2}^{l}=\frac{{-2|t_l|}}{(2\pi a)^{2l+1}} \mathrm{cos}(\eta_{1(n+1)}-\eta_{\bar{1}n}-\varphi +\varphi_0),\label{eq:Htun_eta}
	\end{equation}
	{where $\varphi_0$ is an irrelevant phase shift}.
	In this representation, it is evident that all fields are pinned pairwise, such that the system is indeed fully gapped given that $\mathcal{H}_{T, n+1/2}^{l}$ is the leading relevant term. This can always be achieved for a suitable set of interaction parameters.~\cite{Kane2002} The case $l=0$ corresponds to the IQHE with $\nu=1$, where the gap is opened even without interactions.
	
	We now calculate the FBC in dependence on $\Phi$. At low energies, the argument of the cosine term in Eq.~(\ref{eq:Htun_eta}) is pinned. Taking the sum over $N$, this  implies 
	{$-\sum_n\phi_n(\infty)=N\nu(\varphi-\varphi_0)/2+p\nu\pi$} 
	in terms of the local fields $\phi_n=(\phi_{\bar{1}n}-\phi_{1n})/2$. Here, $p$ is an integer. On the other hand, imposing vanishing boundary conditions at $x=0$ leads to $\sum_n\phi_{n}(0)=-N\pi/2\ \mathrm{mod}\ \pi$. Writing the 2D FBC as $Q_B^{2D}=\sum_n Q_{B,n}^{1D}$, 
	where $Q_{B,n}^{1D}$ is the FBC in the $n$th wire, we get
	{(up to an irrelevant constant)}
	\begin{equation}
	Q_B^{2D}=\frac{\Phi \nu}{2\pi}\frac{e}{\hbar}+{p}\nu\quad\mathrm{mod} \,1.\label{eq:FBC_FQHE}
	\end{equation}
	Thus, the FBC has a linear slope in $\Phi$ which is quantized in units of $\nu e/h$. At fractional filling $\nu=1/(2l+1)$ with $l>0$, this slope is $(2l+1)$ times smaller than in the IQHE case $l=0$.
	Furthermore, there are $2l+1$ different branches of the FBC that cannot be connected under adiabatic evolution of $\Phi$. Finally, the Hall conductance can be obtained from the FBC following Ref.~\onlinecite{Thakurathi2018}, yielding $\sigma_{xy}=e\,\dot Q_B^{2D}/\dot\Phi=e^2\nu/h$ as expected.\cite{note4} Importantly, this result holds for arbitrary changes of $\Phi$ and therefore extends the well-known Laughlin argument, where it is assumed that $\Phi$ is changed by an integer multiple of $h/e$.
	
	\textit{Experimental signatures.} %
	The sample geometry described above can be realized by a Corbino disk,~\cite{Jain2007,Halperin1982,Syphers1986,Fontein1988,Dolgopolov1992,Zhu2017,Schmidt2017} see Fig.~\ref{fig:corbino}. The FBC is then accessible in a rather straightforward way using, e.g., STM techniques \cite{STM1,STM2,STM3,STM4,STM5} to measure the charge located at the boundary of the disk. This allows for several interesting ways to probe the FQHE: Firstly, observing the slope of the linear flux dependence allows one to probe the filling factor, see Eq.~(\ref{eq:FBC_FQHE}). This can further be corroborated by observing the evolution of the FBC as $\Phi$ is varied adiabatically. The FBC will then be $(2\pi/\nu)$-periodic in $\Phi$, with a jump of size unity occurring at a particular value of $\Phi$. Secondly, the different branches of the FBC can be connected if fractional charges are allowed to tunnel between opposite boundaries due to, e.g., a constriction, see again Fig.~\ref{fig:corbino}. By measuring the FBC repeatedly in the presence of a constriction, one finds that it can take $2l+1$ different values, reflecting the $(2l+1)$-fold ground state degeneracy. Similarly, if one now observes the evolution of the FBC with $\Phi$, also jumps of fractional size $s/(2l+1)$, where $s=1,...,2l$ is another integer, can be observed when the system switches from one ground state to another. We note that due to translational invariance it suffices to measure the FBC along a small part of the boundary rather than along the entire circumference, see Fig.~\ref{fig:corbino}. In this case, instead of measuring the absolute values of the slopes and jumps of the FBCs, one should measure their ratios for different filling factors, which again become universal. Thus, boundary charge measurements open up a direct way to probe the fractionalization of charges in the FQHE and, 
	most importantly, allow for a direct experimental verification of the ground state degeneracy. We note that the FBC could alternatively be studied in cold-atom setups with tunable interactions.~\cite{Bernien2017}
	
	\textit{Conclusions.} %
	We studied FBCs in strongly interacting CDW-modulated nanowires and in Corbino disks in the FQHE regime at odd filling factors threaded by an external flux. In both cases, the FBC displays universal features that do not depend on microscopic details of the models such as the exact values of the interaction parameters. In the nanowire (FQHE) case, the FBC depends linearly on the phase offset (flux) with a quantized slope that is determined by the filling factor. Furthermore, the different possible values of the FBC at a fixed phase offset (flux) label different degenerate ground states that cannot be adiabatically connected. The observation of these features is well within experimental reach and opens up a promising route to probe strongly interacting phases via FBCs.
	
	As an outlook, we note that our findings can readily be extended to more general filling factors $\nu=k/(2l+1)$, where $k$ is an integer that is coprime to $2l+1$. A given branch of the FBC will be $2\pi(2l+1)$-periodic under adiabatic evolution of $\Phi$ with $k$ jumps of size unity occurring at specific values of $\Phi$.
	
	\textit{Acknowledgments.} %
	We thank Flavio Ronetti for helpful discussions.
	This work was supported by the Deutsche Forschungsgemeinschaft via RTG 1995, the Swiss National Science Foundation (SNSF) and NCCR QSIT and by the Deutsche Forschungsgemeinschaft (DFG, German Research Foundation) under Germany's Excellence Strategy - Cluster of Excellence Matter and Light for Quantum Computing (ML4Q) EXC 2004/1 - 390534769. We acknowledge support from the Max Planck-New York City Center for Non-Equilibrium Quantum Phenomena. Simulations were performed with computing resources granted by RWTH Aachen University under project thes0753. Funding was received from the European Union's Horizon 2020 research and innovation program (ERC Starting Grant, Grant Agreement No. 757725). 
	
	\appendix
	
	\section{Ground State for Finite and Open System}
	
	When calculating the ground state of the effective one-dimensional model [Eq.~(5) of the main text] for an open and finite system with a fixed number of particles $N_p$, one gets a huge degeneracy due to the missing particles at the system boundaries. To avoid this degeneracy we cut the system and take only $N_s=m Z N_p - (m-1)Z = 9 N_p - 6$ sites instead of $\bar{N}_s=9 N_p$ sites (for $Z=3$ and $m=3$). The degeneracy is then lifted and there is only one possible ground state. This procedure corresponds to forcing the last $6$ sites to be empty. The other two possible ground states that would occur in the thermodynamic limit can then be found by putting either $3$ or all $6$ empty sites to the other boundary of the chain.
	We will use this procedure for our DMRG calculations as well as for the analytical calculations of the boundary charge. 
	
	Using this ground state search, one gets a periodicity of $2\pi$. To get the periodicity of $6\pi$, we calculate all three ground states. We expect these states to evolve into each other when executing an adiabatic time evolution in the grand-canonical ensemble with the chemical potential located in the charge gap. One then gets the periodicity of $6\pi$ which we show in the main text. For convenience, we choose the chemical potential in such a way that the jumps of the adiabatic time evolution and the ones of the ground state search occur at the same position. The positions of the jumps in the adiabatic time evolution may change slightly when changing the chemical potential within the gap.
	
	The average of the FBC is given by
	\begin{align}
	\label{eq:boundary_charge}
	Q_B = \sum_{n=1}^{\bar{N}_s} f_n (\rho_n - N_p/\bar{N}_s),
	\end{align}
	where $\bar{N}_s$ is the number of sites including the $6$ empty sites, $N_p$ is the number of particles, and $\rho_n=\langle a^\dagger_n a_n\rangle$. The envelope function is denoted by $f_n$ and needs to decay smoothly from 1 to 0. For our numerical calculations we take a linear slope for the decay of length $l_p$. The center of this slope has a distance of $L_p$ to the left boundary. For the calculations shown in Fig.~3 of the main text we use $N_s=174$ ($\bar{N}_s=180$) with $L_p=90$ and $l_p=90$.
	
	\section{Analytical Calculation of the FBC}
	In this section we calculate the FBC for the effective one-dimensional model in dependence of the phase $\alpha$ analytically. We focus on the case of $Z=3$ and $m=3$ (other cases can be treated analogously) and consider the atomic limit with strong electron-electron (Coulomb) interaction $U_l=U\gg v_{ex} \gg t$. We introduce an effective unit cell of $Z_{\text{eff}} = Z m = 9$, so that the average bulk density is $\bar{\rho}_B= \frac{1}{Z_{\text{eff}}}=\frac{1}{9}$.
	
\begin{figure}
		\centering
		\includegraphics[width =\columnwidth]{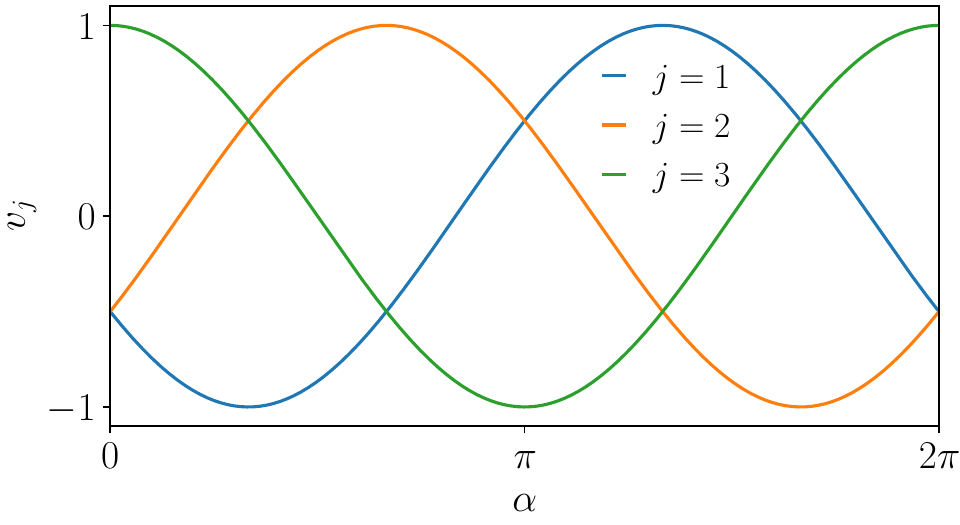}
		\caption{On-site potentials $v_j (\alpha) = \cos (\frac{2 \pi}{3} j+\alpha)$ for $j=1,2,3$.}
		\label{fig:cos_3min}
	\end{figure}

	In the atomic limit the problem of finding the FBC in the given strongly interacting model can be reduced to an effective single-particle model, in which a particle can occupy one of the first $Z$ sites of the effective unit cell with $Z_{\text{eff}}$ sites. As shown in Ref.~\onlinecite{Pletyukhov2020c} the FBC in this limit is dominantly given by the polarization contribution deep in the bulk, which has the form
	\begin{align}
	Q_B \approx Q_P &= - \frac{1}{Z_{\text{eff}}} \sum_{j=1}^{Z_{\text{eff}}} j ( \rho^{\rm bulk}_j - \bar{\rho}_B) \\&= - \frac{1}{Z_{\text{eff}} } \sum_{j=1}^Z j  \rho^{\rm bulk}_j + \frac{Z_{\text{eff}}+1}{2 Z_{\text{eff}}}  \\&=- \frac{1}{9 } \sum_{j=1}^Z j \rho^{\rm bulk}_j + \frac{5}{9}  .
	\end{align}
	Depending on the minima of the cosine potential (see Fig.~\ref{fig:cos_3min}), a particle can sit either on site $j=1$ (for $0<\alpha < \frac{2 \pi}{3}$), or on $j=2$ (for $\frac{4 \pi}{3}<\alpha< 2 \pi$), or on $j=3$ (for $\frac{2 \pi}{3}<\alpha < \frac{4 \pi}{3}$). We thus get three plateaus, 
	\begin{align}
	Q_B \left(0<\alpha < \frac{2 \pi}{3}\right) & \approx - \frac{1}{9 } + \frac59 = \frac49, \\
	Q_{B} \left(\frac{4 \pi}{3}<\alpha< 2 \pi\right) & \approx - \frac{2}{9 }  + \frac59 = \frac{3}{9}, \\
	Q_B \left(\frac{2 \pi}{3}<\alpha < \frac{4 \pi}{3}\right) & \approx - \frac{3}{9 }  +\frac59 = \frac{2}{9}.
	\end{align}
	
	\subsection{Vicinity of $\alpha=0, \frac{2\pi}{3}, \frac{4\pi}{3}$}
	At $\alpha =0, \frac{2 \pi}{3}, \frac{4 \pi}{3}$ the minima contain two sites with the same on-site potential.
	Therefore we consider the first order degenerate perturbation theory in $t$ in the three different intervals around these values. 
	
	1) \underline{$\frac{\pi}{3} < \alpha < \pi$}: $| \psi_0 \rangle \approx c_1^{(1)} |1 \rangle + c_3^{(1)} |3 \rangle $. Then we find
	\begin{align}
	Q_B \left(\frac{\pi}{3} < \alpha < \pi\right) &\approx - \frac19 \left( |c_1^{(1)}|^2 + 3 | c_3^{(1)}|^2 \right) + \frac59 \\ &=  - \frac19 \left(- |c_1^{(1)}|^2 +  | c_3^{(1)}|^2 \right) + \frac39.
	\label{QB1}
	\end{align}
	However, this formula is incorrect, because the hybridization between $|1 \rangle$ and $| 3 \rangle$ is in $O (t)$ impossible, and we need to revise the above result. 
	
	Consider the two subintervals 1a) $\frac{\pi}{3} <\alpha < \frac{2 \pi}{3}$; 1b) $\frac{2 \pi}{3} <\alpha < \pi$.
	
	1a) For $\frac{\pi}{3} <\alpha < \frac{2 \pi}{3}$ the density is mostly located on $|1 \rangle$, with a small admixture of $| 0 \rangle$, which replaces $| 3 \rangle$ in Eq.~\eqref{QB1}. Thus the correct expression reads
	\begin{align}
	Q_B \left(\frac{\pi}{3} < \alpha < \frac{2\pi}{3}\right) &\approx - \frac19 \left( |c_1^{(1)}|^2 + 0 | c_3^{(1)}|^2 \right) + \frac59 \notag\\
	&\approx  - \frac{1}{18} \left( |c_1^{(1)}|^2 -  | c_3^{(1)}|^2 \right) + \frac12.
	\label{QB1a}
	\end{align}
	
	1b) For $\frac{2\pi}{3} <\alpha < \pi$ the density is mostly located on $|3 \rangle$, with a small admixture of $| 4 \rangle$, which replaces $| 1 \rangle$ in Eq.~\eqref{QB1}. Thus the correct expression reads
	\begin{align}
	Q_B \left(\frac{2 \pi}{3} < \alpha < \pi\right) &\approx - \frac19 \left( 4 |c_1^{(1)}|^2 + 3 | c_3^{(1)}|^2 \right) + \frac59 \notag\\
	&\approx  - \frac{1}{18} \left( |c_1^{(1)}|^2 -  | c_3^{(1)}|^2 \right) + \frac16.
	\label{QB1b}
	\end{align}
	Comparing Eqs.~\eqref{QB1a} and \eqref{QB1b} and taking into account that 
	$|c_1^{(1)}|^2 - |c_3^{(1)}|^2$ is a continuous function of $\alpha$ (see below) vanishing at $\alpha = \frac{2 \pi}{3}$, we observe that at this value of $\alpha$ the boundary charge value jumps from $\frac12$ to $\frac16$, such that the jump value is $\frac16 - \frac12 = -\frac13$.
	
	\hspace{0.5cm} 
	
	2) \underline{$\pi < \alpha < \frac{5 \pi}{3}$}: $| \psi_0 \rangle \approx c_3^{(2)} |3 \rangle + c_2^{(2)} |2 \rangle $. This gives us
	\begin{align}
	Q_B \left(\pi < \alpha < \frac{5\pi}{3}\right) &\approx - \frac19 \left( 3 |c_3^{(2)}|^2 + 2 | c_2^{(2)}|^2 \right) + \frac59 \notag\\
	&\approx - \frac{1}{18} \left(  |c_3^{(2)}|^2 - | c_2^{(2)}|^2 \right) + \frac{5}{18}.
	\end{align}
	
	\hspace{0.5cm}
	
	3) \underline{$-\frac{\pi}{3} < \alpha < \frac{\pi}{3}$}: $| \psi_0 \rangle \approx c_2^{(3)} |2 \rangle + c_1^{(3)} |1 \rangle $, leading to 
	\begin{align}
	Q_B \left(-\frac{\pi}{3} < \alpha < \frac{\pi}{3}\right) &\approx - \frac19 \left(2 |c_2^{(3)}|^2 +  | c_1^{(3)}|^2 \right) + \frac59 \notag \\
	&\approx  - \frac{1}{18} \left( |c_2^{(3)}|^2 -  | c_1^{(3)}|^2 \right) +  \frac{7}{18}.
	\end{align}
	
	The coefficients $c_a$ and $c_b$ are found from the eigenvalue problem
	\begin{align}
	&\left( \begin{array}{cc} \frac{v_a - v_b}{2} & - t \\ -t & - \frac{v_a - v_b}{2} \end{array} \right) \left( \begin{array}{c} c_a \\ c_b \end{array} \right) \notag\\&= - \sqrt{\left( \frac{v_a - v_b}{2} \right)^2 + t^2} \left(  \begin{array}{c} c_a \\ c_b \end{array} \right)
	\end{align}
	and it follows
	\begin{align}
	c_b &= \left[  \frac{v_a - v_b}{2} + \sqrt{\left( \frac{v_a - v_b}{2} \right)^2 + t^2}\right] \frac{c_a}{t}, \\
	c_a^2 &=  \frac{t^2}{\left[  \frac{v_a - v_b}{2} + \sqrt{\left( \frac{v_a - v_b}{2} \right)^2 + t^2}\right]^2 +t^2},
	\end{align}
	\begin{align}
	c_a^2 - c_b^2 &= \frac{t^2 - \left[  \frac{v_a - v_b}{2} + \sqrt{\left( \frac{v_a - v_b}{2} \right)^2 + t^2}\right]^2 }{\left[  \frac{v_a - v_b}{2} + \sqrt{\left( \frac{v_a - v_b}{2} \right)^2 + t^2}\right]^2 +t^2} \\&= - (v_a-v_b) \frac{  \frac{v_a - v_b}{2}  +  \sqrt{\left( \frac{v_a - v_b}{2} \right)^2 + t^2} }{\left[  \frac{v_a - v_b}{2} + \sqrt{\left( \frac{v_a - v_b}{2} \right)^2 + t^2}\right]^2 +t^2} .
	\end{align}
	The results are shown in Fig.~\ref{fig:numerics_explanation} for certain parameters, where we also compare them to DMRG data.

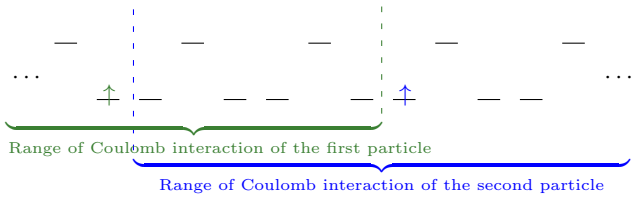
\begin{figure}
		\begin{picture}(11,3.8)
		\label{fig:unused}
		
		\put(0.75,3){\line(1,0){0.4}}
		\put(1.5,2){\line(1,0){0.4}}
		\put(2.25,2){\line(1,0){0.4}}
		\put(3,3){\line(1,0){0.4}}
		\put(3.75,2){\line(1,0){0.4}}
		\put(4.5,2){\line(1,0){0.4}}
		\put(5.25,3){\line(1,0){0.4}}
		\put(6,2){\line(1,0){0.4}}
		\put(6.75,2){\line(1,0){0.4}}
		\put(7.5,3){\line(1,0){0.4}}
		\put(8.25,2){\line(1,0){0.4}}
		\put(9,2){\line(1,0){0.4}}
		\put(9.75,3){\line(1,0){0.4}}
		\put(1.61,1.97){\color{OliveGreen}{$\uparrow$}}
		\put(6.86,1.97){\color{blue}{$\uparrow$}}
		\put(0.0,2.3){$\cdots$}
		\put(10.5,2.3){$\cdots$}
		
		\multiput(2.15,1.1)(0,0.3){9}{\color{blue}{\line(0,1){0.1}}}
		\multiput(6.55,1.75)(0,0.3){7}{\color{OliveGreen}{\line(0,1){0.1}}}
		
		\put(-1.0,1.66){\color{OliveGreen}{$\underbrace{\hspace{5.0cm}}_{\hspace{0.7cm}\text{Range of Coulomb interaction of the first particle}}$}}
		\put(2.15,1.0){\color{blue}{$\underbrace{\hspace{6.6cm}}_{\text{Range of Coulomb interaction of the second particle}}$}}
		\end{picture}
		\caption{States that are ground states at $t=0$ but can be neglected at $t\neq 0$. Two of the particles have only a distance of $(m-1)$ minima. The energy is higher than for the states where every $m$th minimum is occupied as the coupling of the green particle to the right and of the blue one to the left is much smaller in these cases.}
		\label{fig:neglected}
	\end{figure}

	When we made the ansatz that we only need one particle in a cell of $Z_{\text{eff}}=9$ sites, we assumed that 
	there are always two empty minima between the particles due to the repulsive electron-electron interaction. 
	However, it is possible to have configurations where there is only one empty minimum between two particles. 
	Then, both particles need to be located on the outer site of their minimum as shown in Fig.~\ref{fig:neglected}.
	In this case they also do not `see' each other's Coulomb interaction and it would be a ground state for $t=0$. 
	\unitlength0.75cm
	However, we do not need to consider them for the case with $t\neq 0$. Indeed, the neglected states are not coupled to the used ones in the orders that we look at. So there are no neglected couplings.
	Additionally, all states that have a contribution of those new states should have a larger energy than the calculated ground state because the coupling to the adjacent site is much smaller for the configurations shown in Fig.~\ref{fig:neglected} due to the Coulomb interaction.
	Therefore, these states cannot contribute to the ground state for $t\neq 0$.
	
	Using this degenerate perturbation theory, we find some discontinuities at $\alpha = \frac{\pi}{3}, \pi, \frac{5 \pi}{3}$ (see Fig.~\ref{fig:numerics_explanation}) because in the vicinity of these points, there are not two sites in the minimum. In the next section we will remove these discontinuities by treating the vicinities of these points in second order in $t$ with a non-degenerate perturbation theory.

	\subsection{Vicinity of $\alpha=\frac{\pi}{3}, \pi, \frac{5\pi}{3}$}
	In the vicinity of these points we can use non-degenerate perturbation theory where, up to first order in the perturbation, the ground state is given by
	\begin{align}
	\Ket{\Psi}=\Ket{n}+\sum_{m\ne n} \frac{V_{mn}}{E_n-E_m} \Ket{m} \,.
	\end{align}
	Here, $\Ket{n}$ denotes the ground state for $t=0$, and $V_{mn}$ are the matrix elements between the ground state and the excited states $m$, which are given by the hopping in our model. 
	
	In the given regions there is one site in each minimum of the on-site potential. We will call this site $b$, while the two adjacent sites will be called $a$ and $c$. We then get
	\begin{align}
	\Ket{\Psi_0}=\Ket{b}+\frac{t}{v_c - v_b} \Ket{c} +\frac{t}{v_a - v_b} \Ket{a} \,
	\label{eq:GS_pert}
	\end{align}
	for the ground state. Taking into account that this state is not normalized, we get
	\begin{align}
	|c_b|^2&=1-\left(\frac{t}{v_a-v_b}\right)^2-\left(\frac{t}{v_c-v_b}\right)^2 + \mathcal O\left(t^3\right) \, , \\
	|c_a|^2&=\left(\frac{t}{v_a-v_b}\right)^2+ \mathcal O\left(t^3\right), \\
	|c_c|^2&=\left(\frac{t}{v_c-v_b}\right)^2+ \mathcal O\left(t^3\right)\, .
	\end{align}
	
\begin{figure}[t!]
		\centering
		\includegraphics[width =\columnwidth]{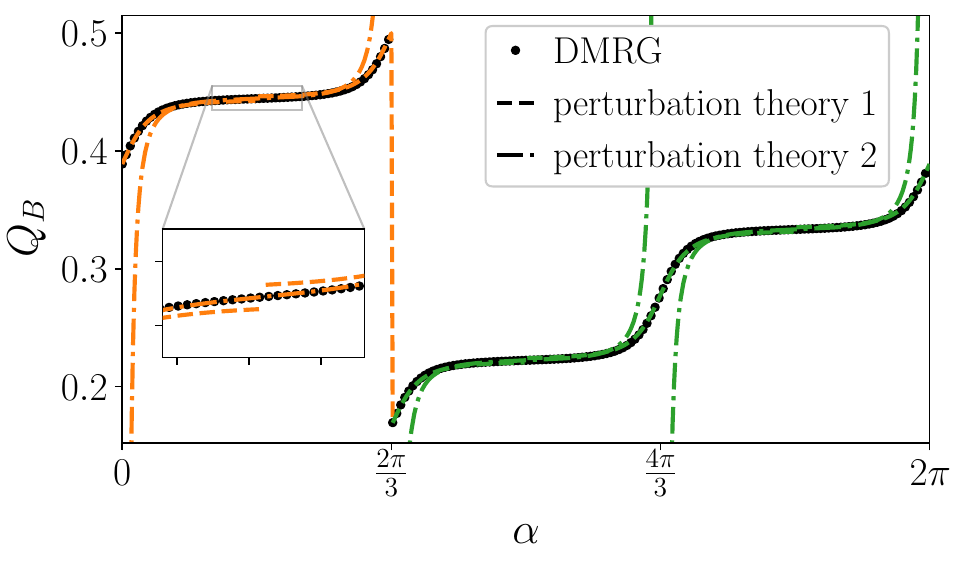}
		\caption{
			Boundary charge as a function of $\alpha$ for the effective model calculated with DMRG and perturbation theory. The other parameters are $m=3$, $Z=3$, $t=1$, $v_{ex}=5$ and $U_l=U=10$ for $l=1,...,6$.  We used $N_s=174$ and $L_p=90$, $l_p=90$ for the envelope function to get the DMRG results. The dashdotted line shows the results calculated in the vicinity of $\alpha=\frac{\pi}{3}, \pi, \frac{5\pi}{3}$, while the dashed line was calculated in the vicinity of $\alpha=0, \frac{2\pi}{3}, \frac{4\pi}{3}$.}
		\label{fig:numerics_explanation}
	\end{figure}

	The boundary charge in the three different regions can then be calculated as follows:
	\begin{align}
	Q_B &\left(0<\alpha < \frac{2 \pi}{3}\right) \notag\\
	& \approx - \frac{1}{9 } \left(0|c_3^{(1)}|^2+|c_1^{(1)}|^2+2|c_2^{(1)}|^2 \right) + \frac59 \notag\\
	&\approx \left(-|c_3^{(1)}|^2+|c_2^{(1)}|^2 \right) +\frac49,
	\end{align}
	\begin{align}
	Q_{B} &\left(\frac{4 \pi}{3}<\alpha< 2 \pi\right) \notag\\
	& \approx - \frac{1}{9 }\left(2|c_2^{(2)}|^2+3|c_3^{(2)}|^2+4|c_1^{(2)}|^2 \right)  + \frac59 \notag\\
	&\approx \left(-|c_2^{(2)}|^2+|c_1^{(2)}|^2 \right) +\frac29, 
	\end{align}
	\begin{align}
	Q_B &\left(\frac{2 \pi}{3}<\alpha < \frac{4 \pi}{3}\right) \notag\\
	& \approx - \frac{1}{9 } \left(|c_1^{(3)}|^2+2|c_2^{(3)}|^2+3|c_3^{(3)}|^2 \right) +\frac59 \notag\\
	&\approx \left(-|c_1^{(3)}|^2+|c_3^{(3)}|^2 \right) +\frac39 \, .
	\end{align}
	We then insert
	\begin{align}
	|c_c|^2-|c_a|^2 \approx \left(\frac{t}{v_c-v_b}\right)^2-\left(\frac{t}{v_a-v_b}\right)^2
	\end{align}
	to get the final result that is shown in Fig.~\ref{fig:numerics_explanation} together with the results of the first order perturbation theory calculated above.
	
	\subsection{Uniting the results}
	In the previous sections we calculated the behavior of the boundary charge in different regimes of $\alpha$.
	To get a final expected curve, one needs to decide where to change between those regimes. Basically we have two different functions for the boundary charge. One should be valid around $\alpha=\pi/3, \pi, 5\pi/3$ and the other one around $\alpha=0, 2\pi/3, 4\pi/3$.
	These two results are plotted in the whole interval of $[0, 2\pi]$ in Fig.~\ref{fig:numerics_explanation}.
	
\begin{figure}
		\centering
		\includegraphics[width =\columnwidth]{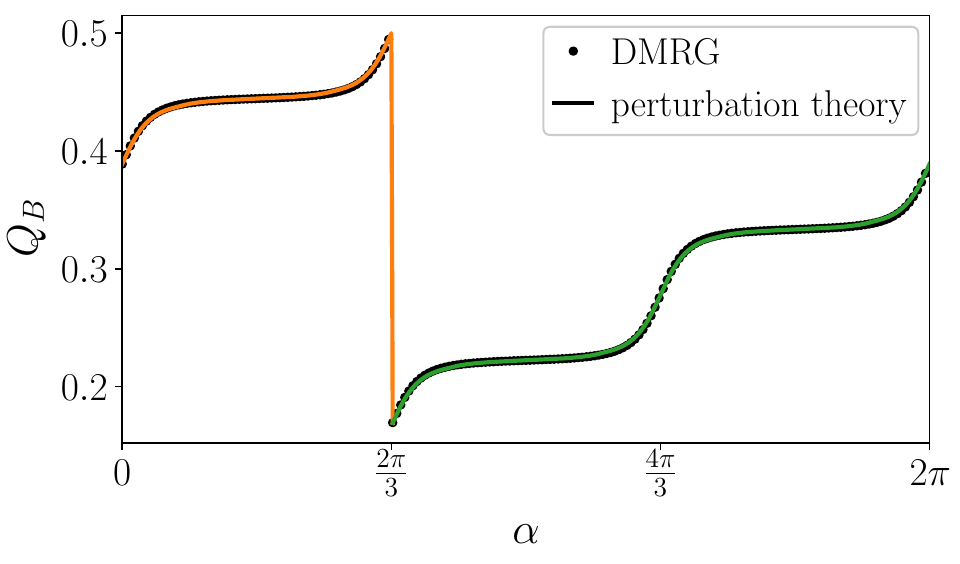}
		\caption{
			Boundary charge as a function of $\alpha$ for the effective model calculated with DMRG and perturbation theory. The other parameters are $m=3$, $Z=3$, $t=1$, $v_{ex}=5$ and $U_l=U=10$ for $l=1,...,6$. We used $N_s=174$ and $L_p=90$, $l_p=90$ for the envelope function to get the DMRG results. The two curves of Fig.~\ref{fig:numerics_explanation} are now combined to one final curve.}
		\label{fig:numerics_final}
	\end{figure}

	As one can see, both results fit a certain part of the numerical curve quite well while there are other parts where they show useless behavior like jumps and divergences. Nevertheless they coincide very well in the intermediate regions between the regimes where they were calculated. To get one final analytical curve the method of calculation was changed at the points where both curves cross each other. The final result can be seen in Fig.~\ref{fig:numerics_final}. The numerical and analytical results lie nearly perfectly on top of each other. Fig.~\ref{fig:numerics_final} corresponds to a zoom into Fig.~3 of the main text, where we show all three ground states with their periodicity of $6\pi$. There, we find a jump of unity for each of the ground states because a particle leaves the system at that point. In Fig.~\ref{fig:numerics_final} we see only a jump by 1/3 because the system changes to another ground state as indicated by the colors. Thereby, all particles are shifted by one minimum to get into the new real ground state of our system. As already mentioned above, the other two ground states can be found by forcing other sites to have zero occupation. For the analytical calculation this means that sites $4,5,6$ or $7,8,9$ are occupied instead of sites $1,2,3$. The boundary charge is then changed by $-1/3$ or $-2/3$.
	
	\subsection{Limits of our perturbation theory}
\begin{figure}[t!]
		\centering
		\includegraphics[width =\columnwidth]{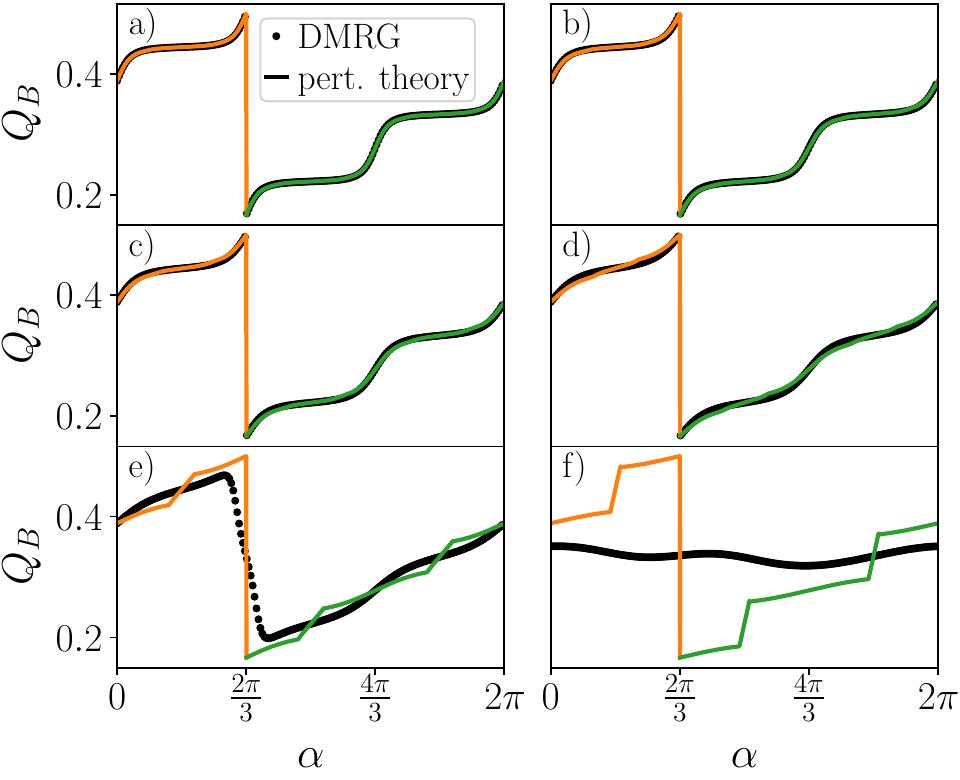}
		\caption{Results of DMRG and perturbation theory for $t=1.0$ and $(v_{ex}, U)=$ (a) $(5.0, 10.0)$, (b) $(4.0, 8.0)$, (c) $(3.0, 6.0)$, (d) $(2.0, 4.0)$, (e) $(1.0, 2.0)$, (f) $(0.5, 1.0)$. The perturbation theory works quite well for large values of $v_{ex}$ and $U$ as expected, while there are clear drawbacks for smaller $v_{ex}$ and $U$. The DMRG results are obtained with $N_s=174$, $L_p=90$, and $l_p=90$.}
		\label{fig:prob_pert}
	\end{figure}

	As we performed the perturbation theory in the regime $U \gg v_{ex} \gg t$, we expect it to fail when $U$ and $v_{ex}$ are not large enough. In Fig.~\ref{fig:prob_pert} we calculate the boundary charge in dependence of $\alpha$ for different $U$ and $v_{ex}$ with constant $U/v_{ex}=2$. 
	For large values of $U$ and $v_{ex}$ the results coincide very well with our numerical DMRG results. For smaller values of $U$ and $v_{ex}$ the curves start to differ. When $U$ and $v_{ex}$ are of the order of $t$ the perturbation theory does not even give us a smooth curve. For those parameters the results of the different regimes of $\alpha$ do not agree in the intermediate region and cannot be united in a satisfying way. 
	
	\section{Dependence on Envelope Function}
\begin{figure}[t!]
		\centering
		\includegraphics[width =\columnwidth]{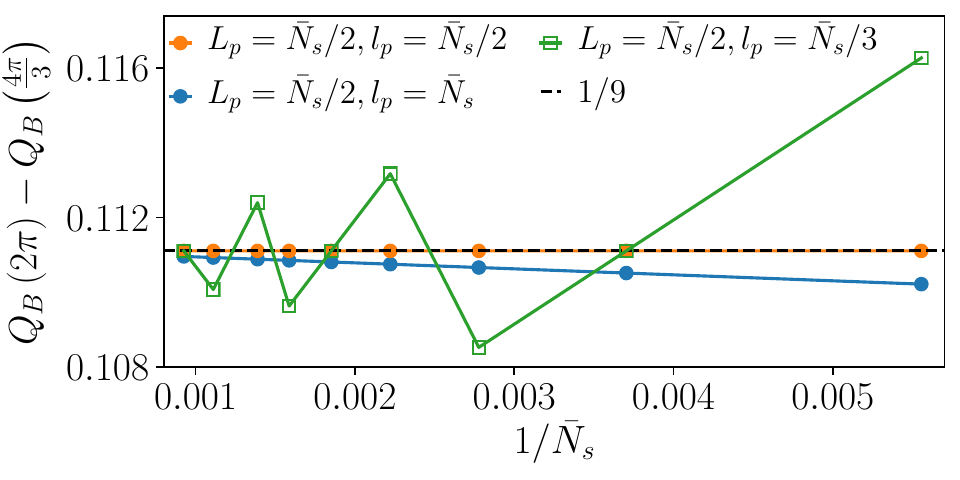}
		\caption{$Q_B\left(2\pi\right)-Q_B\left(\frac{4\pi}{3}\right)$ 
			for different system sizes and different envelope functions. $\bar{N}_s=N_s+(m-1)Z=N_s+6$ describes the system size including the blocked sites where the density is forced to be zero. For $\bar{N}_s\rightarrow\infty$ all curves tend to $1/9$. The other parameters are $t=1$, $v_{ex}=3$, and $U=3$.}
		\label{fig:diff_env}
	\end{figure}

	In the thermodynamic limit the boundary charge needs to be independent of the details of the envelope function. However, the boundary charge can slightly depend on the envelope function for finite system sizes as shown in Fig.~\ref{fig:diff_env}.
	To be as close as possible to the thermodynamic limit, we choose the envelope function with $L_p=\bar{N}_s/2$ and $l_p=\bar{N}_s/2$ (orange curve in Fig.~\ref{fig:diff_env}) for our calculations, where $\bar{N}_s=N_s+6$ denotes the system including the sites that we forced to be empty. With this choice already systems of relatively small size give us a value of $Q_B\left(2\pi\right)-Q_B\left(\frac{4\pi}{3}\right)$ that coincides with the one in the thermodynamic limit.
	
	\section{Disorder}
\begin{figure}[t!]
		\centering
		\includegraphics[width =\columnwidth]{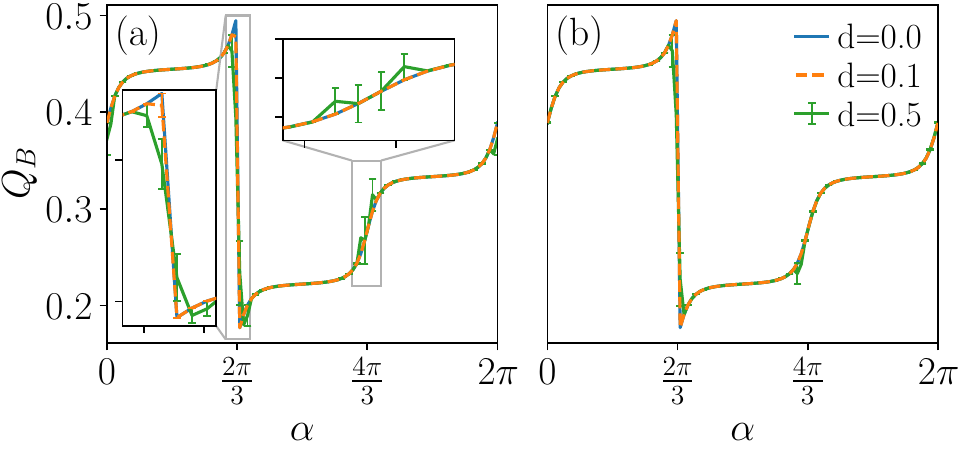}
		\caption{Boundary charge in dependence on $\alpha$ with included (a) hopping and (b) on-site disorder averaged over 20 disorder configurations. The error bars show the size of the uncertainty of the mean value for every second data point in the main panels and for all points in the insets. For $d=0.1$ the error bars are only shown in the left inset as they are smaller than the line width in the other subplots. The other parameters are $m=3$, $Z=3$, $t=1$, $v_{ex}=5$, $U_l=U=10$ for $l=1,...,6$, $N_s=174$, $L_p=90$, and $l_p=90$.}
		\label{fig:disorder}
	\end{figure}

	To prove that our results are stable against disorder, we investigate small random perturbations of the on-site potential and the hopping terms. Therefore, we study the boundary charge in dependence of the phase $\alpha$ averaged over different disorder configurations. The added terms are of the form
	\begin{align}
	- \sum_{n=1}^{N_s-1} w_n (a_n^\dagger a_{n+1} +\mathrm{H.c.}) \text{ and }
	\sum_{n=1}^{N_s} z_n \hat{\rho}_n
	\end{align}
	for hopping and on-site disorder, respectively. Either the parameter $w_n$ or $z_n$ is uniformly distributed in $[-d/2, d/2)$, while the other ones are set to zero.
	The results for a certain set of parameters and 20 disorder configurations are shown in Fig.~\ref{fig:disorder}. For small disorder strengths $d=0.1$ the curve lies on top of the non-disordered one. The quantized slope is therefore stable against small perturbations. Even for stronger disorder $d=0.5$ the average still coincides well with the results of the clean system although there are some differences visible. We see that the jump by $-1/3$ gets smoother because its exact position now depends on the disorder configuration. Therefore the standard deviation  around the mean value of the disorder average is also larger in that region. At transitions between  different plateaus, we also find an enhancement of the standard deviation, while within each plateau the standard deviation is small.
	
	\section{Other Long Ranged Interactions}
\begin{figure}[t!]
		\centering
		\includegraphics[width =\columnwidth]{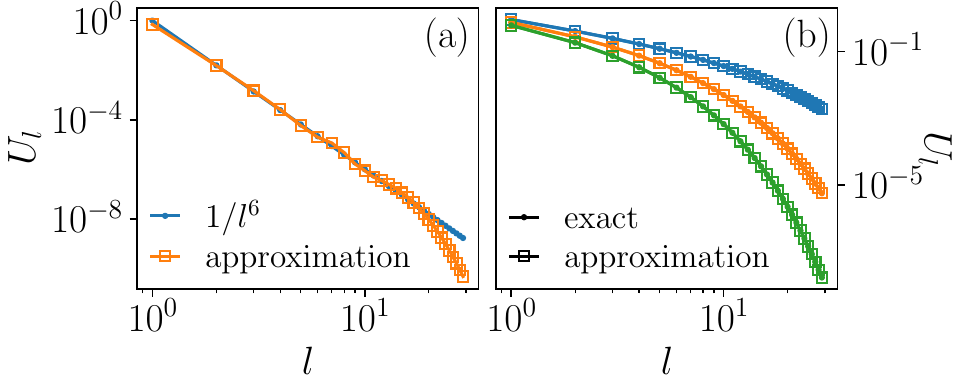}
		\caption{Approximated interaction profile compared to the exact functions (a) $l^{-6}$ and (b) $\exp(-\gamma l)\, l^{-2}$ with $\gamma= 0.1$ (blue), $0.3$ (orange), and $0.5$ (green) on a double logarithmic scale. We use the prefactor (a) $U_0=1$ and (b) $U_0=2$, $N=10$ and $l_{\rm cutoff}=20$ to determine the parameters of the exponential functions.}
		\label{fig:approx}
	\end{figure}

	In the main text and in the previous sections we always considered a constant interaction $U_l=U$ which drops to zero after $l=(m-1)Z$ sites. Here, we will show that we find the same results for other long ranged interactions.
	Thereby, we will investigate a power law of the form $U_l=U_0 \, l^{-6}$ (as experimentally realized in Ref.~\onlinecite{Bernien2017}) and a Yukawa potential of the form $U_l=U_0\, \exp(-\gamma l)\, l^{-2}$. We approximate these long ranged interactions by using a sum of exponential functions ($\sum_{i=1}^N \alpha_i \beta_i^{l}$). The parameters are fitted by minimizing 
	\begin{align}
	\sum_{l=1}^{l_{\rm cutoff}} \left(\ln{\left(\sum_{i=1}^N \alpha_i \beta_i^{l}\right)} - \ln{\left(U_l\right)} \right)^2 \, .
	\end{align}
	Using the  logarithm $\ln$ corresponds to changing the weights of the different data points in the fitting procedure. This assures that the approximation works well even at larger ranges as we know that an interaction is needed on the first $Z(m-1)$ sites. For all functions we only fit the parameters once and rescale them with the prefactor $U_0$. We show the approximations compared to the exact functions in Fig.~\ref{fig:approx}.
	
\begin{figure}[t!]
		\centering
		\includegraphics[width =\columnwidth]{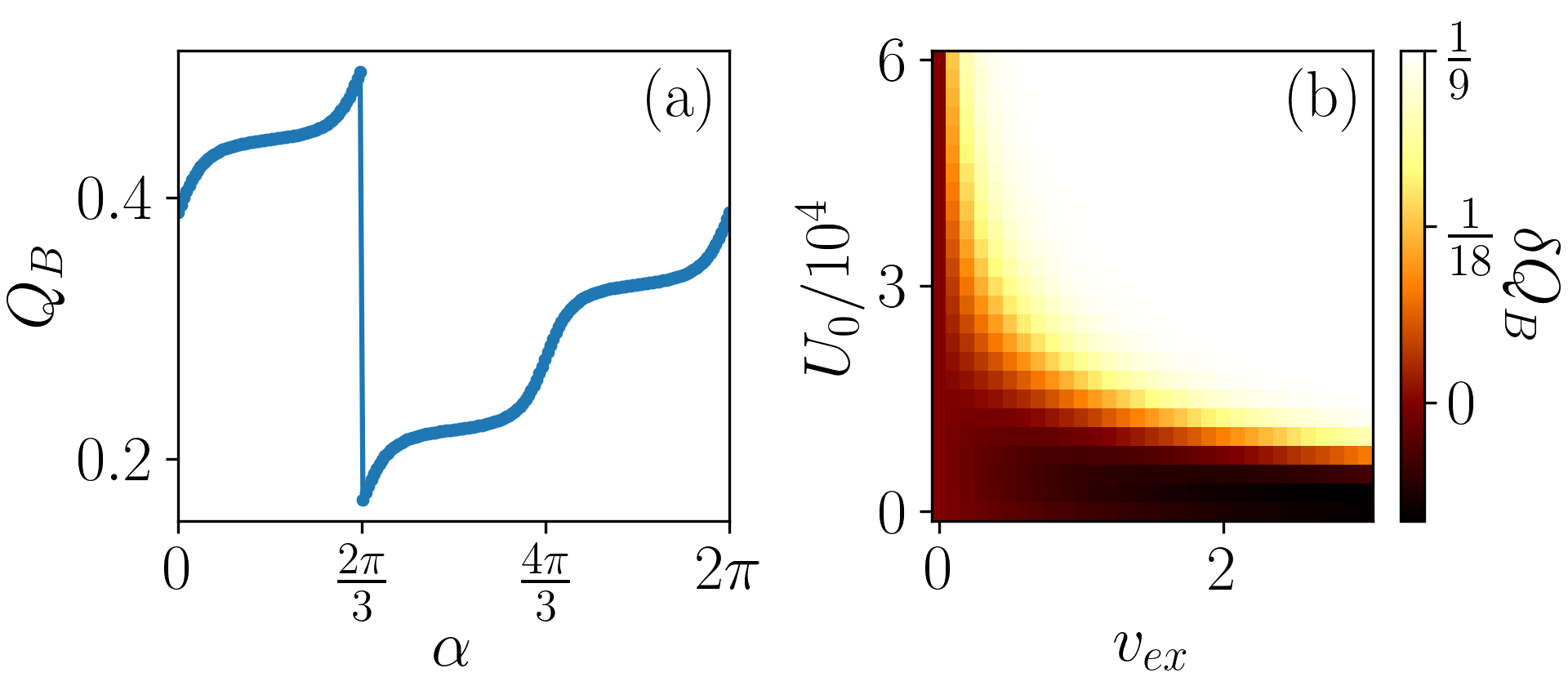}
		\caption{Results for an interaction of $U_l=U_0 \, l^{-6}$ and $Z=m=3$. (a) Boundary charge in dependence on the phase $\alpha$ with $U_0=60000$ and $v_{ex}=3$. (b) False color plot of $\delta Q_B= Q_B \left(2\pi\right) - Q_B\left(\frac{4\pi}{3}\right)$ as a function of $v_{ex}$ and $U$. The fractional slope is stabilized for large $U_0$ and intermediate $v_{ex}$ (white region).}
		\label{fig:powerlaw}
	\end{figure}

	The results of the power law interaction are shown in Fig.~\ref{fig:powerlaw} for $Z=m=3$. With a sufficiently large $U_0$ and $v_{ex}$ we find the fractional slope of $1/2\pi m$ up to a $2\pi/Z$-periodic function as shown in Fig.~\ref{fig:powerlaw}(a). In Fig.~\ref{fig:powerlaw}(b) a false color plot is shown, where the region with $\delta Q_B=1/9$ (white) is where one finds the quantized fractional slope. We find a similar phase boundary compared to the case discussed in the main text. The phase transition occurs at a large prefactor $U_0$ due to the rapid decay of the power law function.
	
\begin{figure}[b!]
		\centering
		\includegraphics[width =\columnwidth]{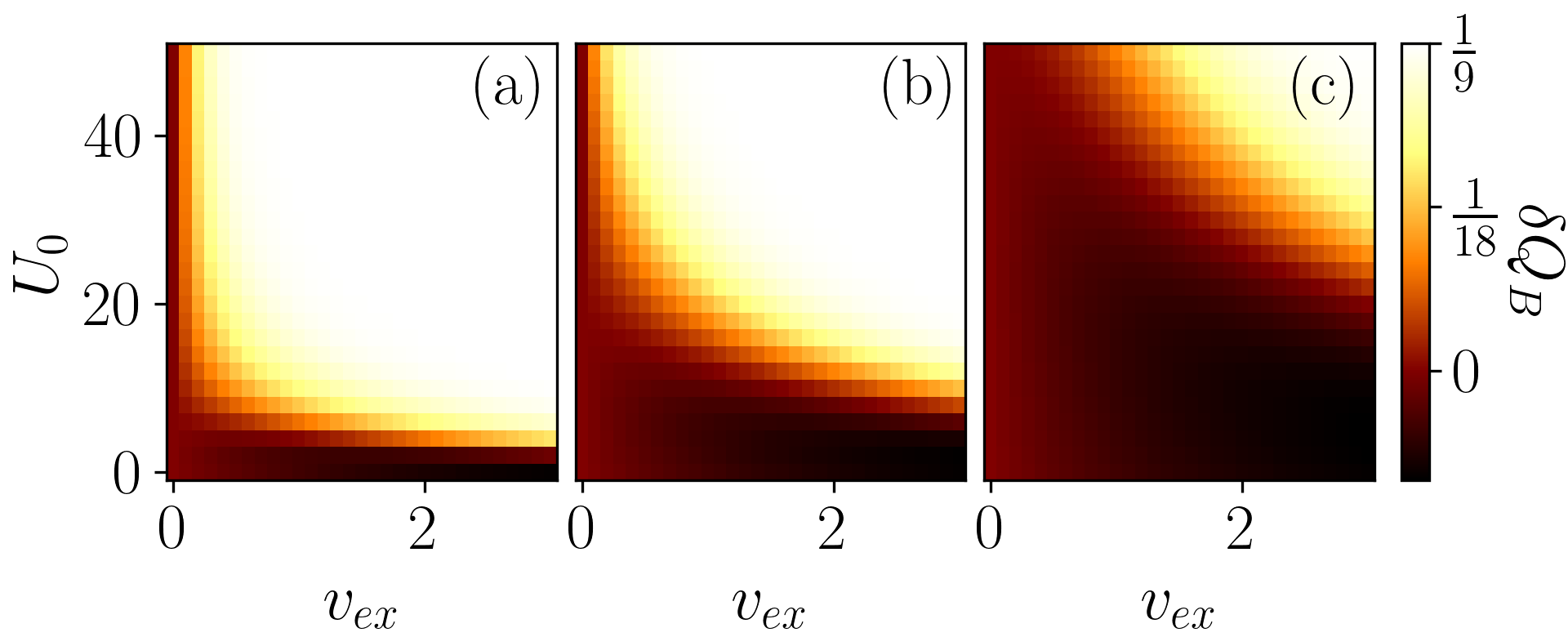}
		\caption{False color plots showing $\delta Q_B= Q_B \left(2\pi\right) - Q_B\left(\frac{4\pi}{3}\right)$ for an interaction of $U_l=U_0\, \exp(-\gamma l)\, l^{-2}$ with (a) $\gamma= 0.1$, (b) $\gamma= 0.3$, and (c) $\gamma= 0.5$. The other parameters are $Z=m=3$. The phase transition is located at larger $U_0$ for larger values of $\gamma$.}
		\label{fig:yukawa}
	\end{figure}

	In Fig.~\ref{fig:yukawa} we show false color plots for the Yukawa potential at different values of the screening $\gamma$. Again, we see a phase transition similar to the cases discussed so far. For larger $\gamma$, the phase boundary is situated at larger $U_0$ because the Yukawa potential decays faster.

	\section{Interface between two CDWs}
	
	In this section, we show how the analytical arguments presented in the main text can be extended to describe the charge located at the interface between two CDWs in a 1D nanowire.
	
	We consider an interface between two CDWs described by a spatially modulated potential of the form
	\begin{align}
	V_m(x)=\begin{cases}
	2 V_m \cos (2mk_Fx+\alpha_<), &x<0,\\
	2 V_m \cos (2mk_Fx+\alpha_>), &x>0,
	\end{cases}
	\end{align}
	where $\alpha_{<}$ ($\alpha_>$) describes the phase offset of the CDW in the domain $x<0$ ($x>0$). We can now follow the same arguments as in the main text for $x\in(-\infty,0)$ and $x\in(0,+\infty)$ separately. In terms of the conjugate bosonic fields $\phi$ and $\theta$, the CDW term then takes the form $H_{CDW}^m=\int dx\,\mathcal{H}_{CDW}^m(x)$ with
	\begin{align}
	\mathcal{H}_{CDW}^m(x)=\begin{cases}
	\frac{{-2|\tilde V_m|}}{(2\pi a)^m} \cos (2m\phi(x)+\alpha_< -\alpha_0),&x<0,\\
	\frac{{-2|\tilde V_m|}}{(2\pi a)^m} \cos (2m\phi(x)+\alpha_> -\alpha_0),&x>0,
	\end{cases}
	\end{align}
	where $\alpha_0$ is again an irrelevant overall phase shift. The CDW term is minimized for the pinning values
	\begin{align}
	\phi(x)=\begin{cases}
	-(\alpha_<-\alpha_0)/2m+l\pi/m,&x<0,\\
	-(\alpha_>-\alpha_0)/2m+n\pi/m,&x>0,
	\end{cases}
	\end{align}
	where $l$ and $n$ are integers. Therefore, the charge located at the interface is given by
	\begin{align}
	Q_D &= - \int^{+\infty}_{-\infty} dx \ \frac{\partial_x \phi (x)}{\pi} \nonumber\\&=- \frac{1}{\pi} [\phi(+\infty)-\phi(-\infty)]\nonumber\\
	&=(l-n)/m+(\alpha_>-\alpha_<)/2\pi m\ \ {\rm mod}\,{1}.
	\label{eq:charge_interface}
	\end{align}
	We thus find that the fractional charge changes linearly with the phase difference $\alpha_>-\alpha_<$ with a slope of $1/2\pi m$.

	Finally, we note that analogous considerations allow us to recover the fractional charge of the excitations in the 2D case. Indeed,
	a bulk excitation in the 2D FQHE corresponds to a kink (domain wall) in the pinned combination of the fields $\eta_{1(n+1)}-\eta_{\bar{1}n}$ for a given $n$, see Eq.~(8) in the main text, while the uniform phase $\varphi$ drops out.
	By using $\sum_n(\eta_{1(n+1)}-\eta_{\bar{1}n})=-2(2l+1)\sum_n\phi_n$ and using that the charge density of a single wire is given by $\rho_n=-\partial_x\phi_n(x)/\pi$ in units of the electron charge $e$, we find that a kink between two adjacent minima of the cosine carries the charge $e/(2l+1)$.
	
\begin{thebibliography}{99}
\bibitem{note1}
		In an experimental realization, it will be more convenient to tune $k_F$ by varying the filling of the wire rather than adjusting the period of the potential.
\bibitem{note6}
		For sufficiently long-ranged interactions we expect our result to also hold when including small perturbations in the filling or the modulation period, leading to small deviations of $m$ or $Z$.
\bibitem{note3}
		Here it is crucial that we work with odd filling factors. At even filling factors, the presence of multiple competing momentum-conserving terms can lead to a more complicated behavior that is not captured here.
\end{thebibliography}
\end{document}